\documentstyle[pra,aps]{revtex}
\begin{document}
\title{Collective ferromagnetism
in two-component Fermi-degenerate gas
trapped in finite potential}
\author{T. Sogo and H. Yabu}
\address{Department of Physics, 
Tokyo Metropolitan University, 
1-1 Minami-Ohsawa, Hachioji, Tokyo 192-0397, Japan}
\date{\today}
\maketitle
\begin{abstract}
Spin asymmetry of the ground states is studied 
for the trapped spin-degenerate (two-component) gases 
of the fermionic atoms 
with the repulsive interaction 
between different components, 
and, for large particle number,
the asymmetric (collective ferromagnetic) states 
are shown to be stable because it can be energetically favorable 
to increase the fermi energy of one component 
rather than the increase of the interaction energy 
between up-down components. 
We formulate the Thomas-Fermi equations 
and show the algebraic methods to solve them.  
From the Thomas-Fermi solutions, 
we find three kinds of ground states in finite system:
1) paramagnetic (spin-symmetric), 2) ferromagnetic 
(equilibrium) and 3) ferromagnetic (nonequilibrium) 
states. 
We show the density profiles and the critical atom numbers 
for these states obtained analytically, 
and,  
in ferromagnetic states, 
the spin-asymmetries are shown to occur
in the central regions of the trapped gas, 
and grows up with increasing particle number. 
Based on the obtained results, 
we discuss the experimental conditions 
and current difficulties 
to realize the ferromagnetic states of the trapped atom gas, 
which should be overcome. 
\end{abstract}
%
%
\section{Introduction}

Recent developments of laser trapping and cooling of atomic gases, 
which realized the Bose-Einstein condensates (BEC) 
for Alkali atoms \cite{RevBEC}, 
have opened up new interests for condensed atomic gases 
with multi-components: 
multi-component BEC \cite{SIS98,BSC01} 
and fermi degenerate (FD) gas \cite{DJ99,GG02,LRT02}, 
and also bose-fermi mixed gas 
in trapped potentials \cite{TSM01,SKC01,GPD02,HSD88}. 

One of interesting physics in such systems 
is the phase structures in the ground states.
In general, multi-component systems show a variety of phase structures 
and phase transitions between them. 
A typical example is seen in the superfluid ${}^3$He 
(a system with $2 \times 2$ components of orbital and spin angular 
momenta), 
where there exist more than two condensed phases resulted 
from different combinations of condensed components \cite{VOL92}. 

In the condensed system of trapped atomic gases
with multi-components, 
several studies have been done 
about new phases and phase transitions: 
the BCS states in attractive two-component fermi gas \cite{HFS97,BP98}, 
a transition to the superfluid states in the crossover region 
between BCS and BEC theories \cite{HKC01,TFM01,OG02}, 
phase separations of different components 
in the BEC \cite{OS98,PB98,GS98}, 
fermi gas \cite{AMM00,SPP00,RF01} and 
bose-fermi mixed system \cite{MOL98,NM99}. 

In this paper, 
we discuss the trapped fermi-degenerate atomic gas 
with two-components: 
$m =\pm 1/2$ (magnetic quantum number) states of 
spin-$1/2$ atoms, or two of the hyperfine states 
of fermionic atoms
with larger total spin ($F > 1/2$). 
Thus, we denote these components as spin degrees of freedom
(up and down components, or up and down atoms) in this paper.

When the interaction between different spin components is weak, 
the stable ground state of the system has a symmetric state
in spin-component distributions at $T=0$ where 
equal number of up and down atoms make similar fermi-degenerate 
distributions so that the total spin of the whole system vanishes 
(paramagnetic state). 
However, 
for sufficiently strong interactions between components, 
an asymmetric state with unequal numbers of up/down atoms can be 
stable because it can be energetically favorable 
to increase the fermi energy of one component 
rather than the increase of the interaction energy 
between up-down components. 
(The interaction energy between up-up/down-down components
can be neglected due to the Pauli blocking effects.) 
We call it ``collective ferromagnetic state'' of two-component 
fermi gas.

As the terms ``paramagnetic'' and ``ferromagnetic'' suggest, 
this theory is related with a mechanism of the metallic magnetism, 
originally proposed by Bloch 
and developed by Stoner et.al. \cite{WG79}, 
where fermi particles are the electrons in conductive bands.  
It should be noted that, in the theory of metallic magnetism, 
a role of $E_{{\rm int}}$ is done 
by the Hartree-Fock exchange energy. 
Studies of ``collective ferromagnetic states'' of atom gas 
are also interesting as experimental testing grounds 
of theoretical ideas in magnetism. 
Recently, such ferromagnetism aroused 
new interests in neutron-star physics as a mechanism of the magnetar 
(neutron stars with strong magnetic fields) \cite{TM00}. 

In the uniform system, 
such ferromagnetic states have been discussed for the atom gas of 
$^6$Li in relation with the stability of the BCS states~\cite{HFS97}. 
The ferromagnetic states have also been discussed 
on the trapped BEC where the interactions between different spin components 
are the origin of the asymmetry \cite{HO98}. 
In this paper, we treat the ferromagnetic states in the trapped 
and finite fermion system. 

In the next section of this paper, 
we formulate a set of equations for the two-component 
fermi gas at $T=0$ and the stability condition 
of its solutions in the Thomas-Fermi (TF) approximations. 
In Section \ref{sec:solution}, solutions of the TF-equations are analyzed 
algebraically, 
and a critical condition for the collective ferromagnetic states 
is obtained. 
In Section \ref{sec:density}, we show the density distributions 
of the two-component fermi gas 
and discuss paramagnetic-ferromagnetic transitions in it. 
In Section \ref{sec:summary}, we summarize the results 
and discuss the experimental conditions 
to obtain the ferromagnetic states.

\section{Thomas-Fermi equations for Two-component Fermi gas}

We consider a $T=0$ system of two-component Fermi gas 
trapped in an isotropic harmonic oscillator potential; 
densities of spin up/down components are denoted by 
$\rho_1(r)$ and $\rho_2(r)$. 
To describe the ground-state behaviors of the system, 
we use the Thomas-Fermi approximation \cite{HFS97,RS80},  
where the total energy of the system is a functional 
of the densities: 
\begin{equation}
     E =\int{d^3r} \left[ 
          \sum_{\sigma=1,2} 
               \left\{
                    \frac{\hbar^2}{2m} 
                    \frac{3}{5} 
                    (6\pi^2)^\frac{2}{3}
                    \rho_\sigma^\frac{5}{3}
         +\frac{1}{2} m \omega^2 r^2 \rho_\sigma 
               \right\}
         +g\rho_1 \rho_2
                   \right].
\label{eQa}
\end{equation}
The $m$ and $\omega$ in (\ref{eQa}) are the fermion mass 
and the oscillator frequency of the trapping potential.
The last term in Eq.~(\ref{eQa}) 
corresponds to the interaction energy 
between different components of fermion, 
and the strength of the coupling constant $g$ 
is given by $g =4 \pi \hbar^2 a /m$ 
where $a$ is the $s$-wave scattering length. 
The interactions between the same components are neglected; 
the elastic $s$-wave scattering is absent 
because of the Pauli blocking effects 
and the $p$-wave scattering is suppressed below $\sim 100$\,{\rm $\mu$K}. 
In the present paper, 
we discuss a system of repulsive interaction, 
so that the parameter $g$ should be positive 
($g>0$).
 
As we show later, 
the collective ferromagnetic states 
of two-component fermi gas show up 
in the case of large particle number ($10^6 \sim 10^{13}$), 
which validates the use of the Thomas-Fermi approximation. 
Also the validity of the approximation can be estimated from 
the smoothness of the mean-field potential: 
$V_{{\rm eff}} =\frac{1}{2} m \omega^2 r^2 +g \rho_{1,2}$. 
As a parameter that represent its smoothness, 
we can take a local de~Broglie wave length 
$\lambda(r) =\hbar/p(r)$, 
where $p(r)$ is a TF local momentum 
defined by $p(r) =\sqrt{2m(\epsilon_F-V_{\rm eff})}$ 
($\epsilon_F$ : Fermi energy). 
As presented in \cite{LL}, 
a validity condition with $\lambda(r)$ is given by 
$f(r) \equiv |\frac{d\lambda}{dr}| \ll 1$. 
Evaluating $f(r)$ with the parameters 
given in the last section (for ${}^{40}$K), 
we obtain $f(r) \lesssim 10^{-3}$ 
except the classical turning points. 
It also supports the validity of the TF approximations 
in the present case.

To simplify Eq.~(\ref{eQa}), 
we introduce the scaled dimensionless variables:
\begin{equation}
     n_\sigma =\frac{128}{9\pi} \rho_\sigma a^3, \quad
     x        =\frac{4}{3\pi} \frac{ar}{\xi^2}, \quad
     {\tilde E} =\frac{2^{18}}{3^7 \pi^6}
                 \left( \frac{a}{\xi} \right)^8
                 \frac{E}{\hbar\omega}, 
\label{eQb}
\end{equation}
where $\xi =\sqrt{\hbar/m\omega}$ is the oscillator length. 
Using these variables, 
Eq.~(\ref{eQa}) becomes 
\begin{equation}
     {\tilde E} =\int{d^3x} \left[ 
                 \sum_{\sigma=1,2} 
                 \left( \frac{3}{5} 
                        n_\sigma^{\frac{5}{3}} 
                       +x^2 n_\sigma 
                 \right) 
                 +n_1 n_2
                            \right].
\label{eQc}
\end{equation}

The Thomas-Fermi equations for the densities $n_{1,2}$ 
are derived from the variations of 
the total Energy $\tilde E$ on $n_{1,2}$
with a constraint on the total particle number ${\tilde N}$:
$\frac{\delta}{\delta n_\sigma} 
({\tilde E} - \lambda {\tilde N} ) =0$,
where ${\tilde N}$ is the scaled total particle number 
defined by 
\begin{equation}
     {\tilde N} 
    ={\tilde N_1} +{\tilde N_2} 
    =\sum_\sigma \int{d^3x} n_\sigma 
    =\frac{2^{13}}{3^5\pi^4}\left( \frac{a}{\xi} \right)^6 N
    \sim 0.346 \left( \frac{a}{\xi} \right)^6 N.
\label{eQe}
\end{equation}
The Lagrange multiplier $\lambda$ in the variational equation is 
for the fermion-number constraint, and 
related with the scaled chemical potential ${\tilde \mu}$ 
through the relation ${\tilde \mu} N =\lambda {\tilde N}$; 
using Eq.~(\ref{eQe}), we obtain
\begin{equation}
     {\tilde \mu} \sim 0.346 \left( \frac{a}{\xi} \right)^6 \lambda.
\label{eQeA}
\end{equation}

It should be noted that, 
in Eq.~(\ref{eQc}),  
parameters $(m, \omega, g)$ has been scaled out 
and no parameters are included except $\lambda$. 
The Lagrange multiplier $\lambda$ is determined 
by the total fermion number ${\tilde N}$, so that ${\tilde N}$  
is the only parameter that determine the ground-state properties 
of the system. 

Using Eqs.~(\ref{eQc}) and (\ref{eQe}) 
for the variational equation, 
we obtain the TF equations: 
\begin{equation}
     n_1^{\frac{2}{3}} +n_2 =\lambda-x^2 \equiv M(x), \quad
     n_2^{\frac{2}{3}} +n_1 = M(x). 
\label{eQf}
\end{equation}

The stability condition for solutions of Eq.~(\ref{eQf}) 
can be derived from the second-order variations 
of the energy functional:
\begin{equation}
     \left|\begin{array}{cc}
           \frac{\delta^2 \tilde E}{\delta n_1^2} & 
           \frac{\delta^2 \tilde E}{\delta n_1 \delta n_2} \\
           \frac{\delta^2 \tilde E}{\delta n_2 \delta n_1} &
           \frac{\delta^2 \tilde E}{\delta n_2^2}
           \end{array}
    \right| \geq 0.
\label{eQg}
\end{equation}
Using Eq.~(\ref{eQc}), 
we obtain the stability condition for the present case:
\begin{equation}
     n_1 n_2 \leq \left( \frac{2}{3} \right)^6.
\label{eQh}
\end{equation}

\section{Solutions of Thomas-Fermi Equations}
\label{sec:solution}

In this section, we show solutions of 
coupled TF equations (\ref{eQf}) in algebraic form 
and discuss their stability 
based on the stability condition (\ref{eQh}). 

Let's introduce variables 
$s$ and $t$ as $n_1 =s^3$ and $n_2 =t^3$, 
then Eq.~(\ref{eQf}) becomes 
\begin{equation}
     s^3 +t^2 =M,  \quad
     t^3 +s^2 =M. 
\label{eQi}
\end{equation}
Making sum and difference of them, 
we obtain two equations equivalent with Eq.~(\ref{eQi}): 
\begin{equation}
     s^3 +t^3 +s^2 +t^2 =2M,  \quad
     s^3-t^3 -(s^2-t^2) =(s-t) (s^2 +st +t^2 -s-t) =0. 
\label{eQj}
\end{equation}
The factorized form of the second equation 
gives two alternatives: 
1) $s=t$ or 2) $s^2+st+t^2 -s-t =0$. 

In the case 1), $s=t$, 
the first equation in (\ref{eQj}) 
becomes $s^3 +s^2 =M$, 
which can be solved algebraically 
with the Caldano's formula. 
It includes only one positive root when $M \geq 0$: 
\begin{equation}
     s =\frac{f(M)}{6} +\frac{2}{3 f(M)} -\frac{1}{3}, \quad
     f(M) =\left[ -8 +108 M +12 \sqrt{81 M^2 -12 M} \right]^{1/3}. 
\label{eQjA}
\end{equation}
For this solution, 
the stability condition (\ref{eQh}) gives
$s \leq 2/3$, 
which leads to the constraint for $M$: 
\begin{equation}
     M =s^3 +s^2 
       \leq \left( \frac{2}{3}\right)^2 
                  +\left( \frac{2}{3} \right)^3 
       =\frac{20}{27}. 
\label{eQk}
\end{equation}
Thus, the stable symmetric solutions exist 
when $0 \leq M \leq 20/27$. 

Next, we take the case 2) :
$s^2+st+t^2 -s-t =0$. 
Now, this and the another equation in (\ref{eQj}) are symmetric 
under the exchange between $s$ and $t$, 
so that they can be represented by 
elementary symmetric polynomials, 
$A \equiv s+t$ and $B \equiv st$:
\begin{equation}
     A^3 -A^2 -A +M =0,  \quad
     A^2 -A -B =0.
\label{eQl}
\end{equation}
The first equation has a real positive solution: 
\begin{equation}
     A =\frac{g(M)}{6} +\frac{8}{3 g(M)} +\frac{1}{3},  \quad
     g(M) =\left[ -108 M +44 +12 \sqrt{81M^2 -66M -15} \right]^{1/3}.
\label{eQm}
\end{equation}
Substituting it in (\ref{eQl}), 
we obtain the solution for $B$:
\begin{equation}
     B =\frac{[g(M)]^2}{36} -\frac{g(M)}{18} 
       +\frac{4}{9} -\frac{8}{9 g(M)} +\frac{64}{9 [g(M)]^2}. 
\label{eQn}
\end{equation}
The $s$ and $t$ can be recovered 
as two solutions of the equation: 
$x^2 -A x +B =0$. 
These solutions are generally asymmetric ($s \neq t$). 

We can check the stability of the asymmetric solutions  
with Eq.~(\ref{eQh}), 
which gives 
a constraint for $B$: 
$B =st =(n_1 n_1)^{1/3} \leq 4/9$. 
Then, from the second equation in (\ref{eQl}), 
we obtain that for $A$: 
$1 \leq A \leq 4/3$. 
Differentiating the first equation in (\ref{eQl}) by $A$, 
we find that $M$ is a monotonically decreasing function 
of $A$ within the interval $1 \leq A \leq 4/3$. 
Combining these results, 
we obtain the stability range of $M$ 
for the asymmetric solutions: $20/27 < M \leq 1$. 

In summary, 
the stable solutions of Eq.~(\ref{eQj}) are
\begin{equation}
\begin{array}{rll}
     1. & 0 \leq M \leq \frac{20}{27} & \mbox{symmetric},  \cr
     2. & \frac{20}{27} < M \leq 1 & \mbox{asymmetric}, \cr
     3. & 1 < M                    & \mbox{(no stable solutions)}. \cr
\end{array}
\label{eQp}
\end{equation}

In case that $1 < M$, Eq.~(\ref{eQj}) have no stable solutions. 
In the TF variational equation 
$\frac{\delta}{\delta n_\sigma} 
({\tilde E} - \lambda {\tilde N} ) =0$, 
we assumed the chemical equilibrium between fermion 1 and 2, 
and put their chemical potentials 
(Lagrange multipliers) equal: $\lambda =\lambda_1 =\lambda_2$. 
Accordingly, in $1 < M$, 
we should say that 
the system has no stable solutions in ``equilibrium''. 
Thus, for $1 < M$, we should take the nonequilibrium states 
where all fermions occupy one component (complete asymmetry). 
If we assume $s=0$, 
$t$ is obtained by solving the equation: 
\begin{equation} 
     t^2 =M,  \quad  s=0.
\label{eQq}
\end{equation}

\section{Density Profiles of Two-component Fermi Gas}
\label{sec:density}

Using the results in the previous section, 
we can calculate the density profiles 
of the two-component fermi gas for any values of $\lambda$. 
Corresponding to the classification by $M =\lambda-x^2$ 
in (\ref{eQp}), 
we should divide $\lambda$ into three regions: 
\begin{equation}
     \hbox{(i) }\, 0 \leq \lambda \leq \frac{20}{27},   \quad
     \hbox{(ii) }\, \frac{20}{27} < \lambda \leq 1,  \quad
     \hbox{(iii) }\, 1 < \lambda,
\label{eQr}
\end{equation}
and discuss qualitative profiles of the density distributions 
$n_{1,2}$. 

\begin{enumerate}
\item[(i)] {\it Paramagnetic ground states.} 
In these cases, 
$M(x)$ satisfies $M(x) =\lambda -x^2 \leq 20/27$ 
for any value of $x$. 
Thus, from (\ref{eQp}), 
the density distributions are composed 
of the symmetric solution (\ref{eQjA})  
in all spatial region. 
Consequently, the ground states are paramagnetic 
in the region (i): $n_1(x) =n_2(x)$. 
In Fig.~\ref{fig:density}(a), the density profiles are shown when $\lambda =1/2$. 
They monotonically decreases with the scaled radial distance $x$, 
and vanishes at the TF cut-off: 
$x_{{\rm TF}} \equiv \sqrt{\lambda} =1/\sqrt{2}$ in this case.
These states occurs in the case of small fermion number.
\item[(ii)] {\it Ferromagnetic ground states in equilibrium.}
In these cases, 
we obtain $M(x) \leq 20/27$ in the outside region 
($x \geq \sqrt{\lambda-20/27}$), but $20/27 < M(x) \leq 1$ 
in the inside region ($x < \sqrt{\lambda-20/27}$). 
Correspondingly, 
the density distributions become symmetric in the outside region 
and asymmetric in the inside region, 
so that the ground states become ferromagnetic in the region (ii). 
Because the condition $M(x) \leq 1$ is satisfied, 
solutions are in equilibrium in all spatial regions, 
and densities are partially asymmetric in the inside region. 
In Fig.~\ref{fig:density}(b), the density profiles are shown when $\lambda =4/5$ 
for $n_1$ (solid line) and $n_2$ (dotted line). 
The border between symmetric and asymmetric regions is given by 
$x_{{\rm AS}} \equiv \sqrt{\lambda-20/27} =\sqrt{8/135}$ 
in this case. 
In the outside (symmetric) region, 
the both lines overlap and vanishes 
at $x_{{\rm TF}} =2/\sqrt{5}$.  
\item[(iii)] {\it Ferromagnetic ground states in nonequilibrium.}
In these cases, 
density profiles are also asymmetric 
in the inside region ($x < x_{{\rm AS}}$) as in the above region. 
However, in the most inside region ($x < \sqrt{\lambda-1}$), 
$M(x) >1$ is satisfied 
and, according to the condition 3 in (\ref{eQp}), 
the density profiles become nonequilibrium and are given by 
complete asymmetric solutions in (\ref{eQq}). 
In Fig.~\ref{fig:density}(c), we show the density profiles when $\lambda=3/2$ 
for $n_1$ (solid line) and $n_2$ (dotted line). 
In the most inside region 
($x < x_{{\rm EQ}} =\sqrt{\lambda-1} =1/\sqrt{2}$), 
the density profiles are completely asymmetric ($n_2 =0$), 
and, in $x_{{\rm EQ}} \leq x < x_{{\rm AS}} =\sqrt{41/54}$, 
they are partially asymmetric. 
In the outside (symmetric) region ($x > x_{{\rm AS}}$), 
two lines overlap and vanishes at $x=x_{{\rm TF}} =\sqrt{3/2}$.
Thus, the ground states are ferromagnetic and in nonequilibrium 
in the region (iii) which corresponds to the cases of large fermion number.
\end{enumerate}

In Fig.~\ref{fig:chemi}, we show the dependence of $\lambda$ 
on the fermion number ${\tilde N}$, 
which is calculated by Eq.~(\ref{eQe}) 
using the fermion densities $n_{1,2}(x)$ obtained above;
for the ground states (solid line) 
and the paramagnetic states (dotted line). 
From the above discussions, 
for $\lambda \leq 20/27$, the ground states are paramagnetic, 
so that the both lines overlap. 
As can be read off in this figure, 
the critical value $\lambda =20/27$ corresponds to ${\tilde N}_C =0.53$, 
which is the critical fermion number 
for the transition between paramagnetic and ferromagnetic states.

In Fig.~\ref{fig:energy}(a), the variations
of the scaled total energies ${\tilde E}$ 
are shown against ${\tilde N}$, 
for the ground (solid line) and paramagnetic (dotted) states. 
The ${\tilde E}$ can be obtained by Eq.~(\ref{eQc}) 
as a function of $\lambda$
using the fermion densities $n_{1,2}(x)$; combined with the ${\tilde N}$-dependence of $\lambda$ 
(shown in Fig.~\ref{fig:chemi}), 
we can obtain its ${\tilde N}$-dependence. 
In ${\tilde N} \leq {\tilde N}_C =0.53$, 
both lines overlap completely because the ground states become 
paramagnetic. 
In Fig.~\ref{fig:energy}(b), we plot the energy difference 
between the ground and paramagnetic states. 
In this figure, 
we can find that, in ${\tilde N} > {\tilde N}_C$, 
the ground-state energy (solid line) shifts lower than 
that of the paramagnetic states (dotted line); 
it shows that the ferromagnetic ground states 
become more stable in this region.

As can be seen in Fig.~\ref{fig:energy}(a) when ${\tilde N} >0.53$, 
the energy difference between ferromagnetic and 
paramagnetic states $|{\tilde E} -{\tilde E}_{{\rm para}}|$ 
is very small in comparison with the total energy 
${\tilde E}$: 
e.g. $|{\tilde E} -{\tilde E}_{{\rm para}}|/{\tilde E} \sim 0.01$ 
at ${\tilde N} =2.0$.
A large part of ${\tilde E}$ consist of the stacked kinetic energy 
due to the fermi degeneracy of fermions. 
Roughly speaking, 
its size can be estimated from the total energy 
of the noninteracting fermi gas ($g=0$) 
trapped in the same harmonic oscillator potential; 
${\tilde E}_{\text{non}}=\frac{3}{16}\pi^2
(\frac{4}{\pi^2}{\tilde N})^{4/3}$,
which increases rapidly with increasing ${\tilde N}$.
The ${\tilde E}_{\text{non}}$ is plotted 
also in Fig.~\ref{fig:energy}(a) 
(dot-dashed line), 
the amounts of which are almost $\sim 75\%$ of ${\tilde E}$. 

However, to evaluate the scale of the energy difference, 
we should compare it with an one-particle excitation energy 
at the Fermi surface, 
which can be estimated from the scaled chemical potential 
${\tilde \mu}$ defined in (\ref{eQeA}). 
In case of ${}^{40}$K  
with the harmonic oscillator frequency 
$\omega=1000\,{\rm Hz}$ 
(for other parameters, see the next section), 
Eq.~(\ref{eQeA}) becomes ${\tilde \mu} \sim 10^{-13} \lambda$. 
When ${\tilde N} \sim 2$ ($\lambda \sim 1$ 
from Fig.~\ref{fig:chemi}), 
the ratio $\Delta{\tilde E}/{\tilde \mu}$ 
becomes $10^{13}$. 
Thus, we should say that the energy difference 
between ferromagnetic and 
paramagnetic states are fairly large. 

In Fig.~\ref{fig:asymmetry}, we show the fermion number asymmetry 
$({\tilde N}_1 -{\tilde N_2})/{\tilde N} 
     ={(N_1 -N_2)/(N_1+N_2)}$ 
against the total fermion number ${\tilde N}$.

\section{Summary and Discussions}
\label{sec:summary}

We discussed the possibility of transition to 
ferromagnetic states in the two-component fermi gas 
using the Thomas-Fermi approximation 
from theoretical points of view. 
Based on the results that we obtained, 
let's discuss experimental conditions 
and also difficulties to observe ferromagnetic ground states 
in the trapped atomic gas. 
We hope that these difficulties are overcome 
in future developments in experimental technics.

As shown in the previous section, 
the ferromagnetic ground states become stable 
when ${\tilde N} \gtrsim 0.53$; 
using Eq.~(\ref{eQe}), 
the unscaled critical number $N_C$ becomes 
\begin{equation}
     N_C \sim {0.53 \over 0.346} \left( \frac{\xi}{a} \right)^6
         =1.5 \left( \frac{\xi}{a} \right)^6. 
\label{eQu}
\end{equation}
As an example, we take the ${}^{40}$K atoms 
(mass $m=0.649\times 10^{-25}\,{\rm kg}$)
trapped in the harmonic oscillator potential 
with $\omega=1000\,{\rm Hz}$. 
For the scattering length, we take the value 
$a=169 a_B$ ($a_B$: Bohr radius) given in \cite{WNG00}.
Using these parameters, we obtain 
$N_C \sim 10^{13}$. 
In recent experiments, the trapped Fermi-degenerate gas 
has been performed up to $\sim 10^6$ atoms, 
so that the larger trapping potential are necessary 
for realization of the ferromagnetic ground state 
than currently used one. 
In addition, because of the high central density 
($\sim 10^{17}\,{\rm cm^{-3}}$) for $\sim 10^{13}$ atoms, 
the inelastic/multi-body scattering processes 
in them becomes important, which might destroy the trapped atoms 
before they reach the required density.

There can be several possibilities for the reduction of $N_C$. 
For example, if the scattering length can be increased 
by the Feshbach resonance for ${}^{40}$K, 
which has been observed experimentally \cite{LRT02},
the value of $N_C$ decreases  
and the ferromagnetic ground states can be obtained 
in small fermion number: 
simple estimation gives, $a=820a_B$ for $N_C \sim 10^9$
and $a=2600a_B$ for $N_C \sim 10^6$. 
However, in current experiments, 
the tuning into Feshbach resonances is done 
by applying the magnetic field, 
which should make the energy difference 
between the up and down states 
and makes the ferromagnetic transition 
into a crossover in the spin asymmetry.
The ways to tune Feshbach resonances 
by non-magnetic external forces (e.g. electric fields) 
are preferred for the present purpose. 

The use of heavier elements, e.g. Sr or Yb, 
is also effective for the ferromagnetic states. 
We hope that the combination of these methods 
may lead to the experimental achievements. 

We also comment on the process to perform the ferromagnetic states. 
There exist two possibilities: 
\begin{enumerate}
\item[a)]  The experiment starts with the magnetic field 
in some direction which makes the spin asymmetry (e.g. $N_1 > N_2$). 
Then, the magnetic field is switched off adiabatically 
in the cooling process. When the number of the remaining atoms 
is enough large, it shows the ferromagnetic states.
\item[b)] The experiment starts with the symmetric trap 
($N_1 =N_2$), and, after some relaxation time elapses, 
the atoms release their spin angular momenta 
and become the ferromagnetic state. 
\end{enumerate}
The case a) is similar with a standard processe 
for observing the phase transition in the ferromagnetic materials.
In case b), the spin relaxation time is considered to be the same 
order with that of clusterization, 
so that it might be difficult to observe the ferromagnetic transition 
within the lifetime of the atomic gas.

Finally, we comment on the spacially phase-separated states 
in the trapped fermionic gas 
in the case of the large particle number and interaction 
strength~\cite{AMM00,SPP00,RF01}. 
They correspond to the phase separation in the uniform system 
(two-phase coexistent region), discussed in \cite{HFS97}.
The naive TF calculations give smaller energy 
for the ferromagnetic states, but the energy difference 
is very small. 
The competition/coexistence of these states 
should be an interesting problem in future. 

We are very grateful to T. Maruyama, T. Suzuki and T. Miyakwa 
for fruitful discussions. 
Special thanks are also for T. Tatsumi 
for introducing us the physics of collective ferromagnetism.



\begin{figure}
\caption{\label{fig:density}
Density profiles of fermions, 
$n_1$ (solid line) and $n_2$ (dotted line) 
for
(a) $\lambda =0.5$, (b) $0.8$ and (c) $1.5$. 
The scaled densities $n_{1,2}$ and the scaled radial distance $x$ 
are dimensionless and defined by Eq.~(\ref{eQb}). 
The $\lambda$ is a (dimensionless) Lagrange multiplier.
The (a)-(c) correspond to 
a paramagnetic state (i), 
a ferromagnetic state in equilibrium (ii), 
a ferromagnetic state in nonequilibrium (iii) 
each other. 
The scaled fermion number ${\tilde N}_{1,2}$ 
are
$\tilde N_1=\tilde N_2=0.089$ for (a),
$({\tilde N}_1,{\tilde N}_2)= (0.33,0.31)$ for (b), 
and $(3.50,0.68)$ for (c).}
\end{figure}

\begin{figure}
\caption{\label{fig:chemi}
Lagrange multiplier $\lambda$ against 
the scaled fermion number ${\tilde N}$. 
The solid and dotted lines are 
for the ground and paramagnetic states. 
The $\lambda$ is introduced as a Lagrange multiplier 
for fermion-number constraint, and the ${\tilde N}$ is defined by 
Eq.~(\ref{eQe}). 
Both quantities are dimensionless.}
\end{figure}

\begin{figure}
\caption{\label{fig:energy}
(a) The scaled total energy ${\tilde E}$ 
against the scaled total fermion number 
${\tilde N} ={\tilde N}_1 +{\tilde N}_2$: 
the solid line is for the ground states 
and the dotted one for the paramagnetic states. 
The dash-dotted line is for the total energy of the 
noninteracting fermionic system, 
$\tilde E_{\rm non}=\frac{3}{16}\pi^2
(\frac{4}{\pi^2}{\tilde N})^{4/3}$. 
(b) The energy differences from that of the paramagnetic states,  
${\tilde E}-{\tilde E}_{{\rm para}}$. 
The solid and dotted lines are 
for the ground and paramagnetic states.}
\end{figure}

\begin{figure}
\caption{\label{fig:asymmetry}
Variation of the fermion number asymmetry 
$({\tilde N}_1 -{\tilde N}_2)/{\tilde N}$ 
against the scaled total fermion number ${\tilde N}$. 
The solid and dotted lines are 
for the ground and paramagnetic states each other.}
\end{figure}
\end{document}